# *C*-symmetric quantization of fields leading to a natural normal ordering

Zahid Zakir[*)]

*Centre for Theoretical Physics and Astrophysics,*

*11a, Sayram, 100170 Tashkent, Uzbekistan*

**Abstract**

At the quantization of fields, due to the non-linear character of the time reversal, the creation-annihilation operators for the negative frequency modes should be replaced to the operators of antiparticles not directly in the field operators, but in the operator products. For the standard minimal Lagrangians (asymmetrical under the complex conjugated fields) it is shown that the charge conjugation (C-) symmetry conditions for the Hamiltonian and the charge operator lead to the identities for the operator products allowing one to replace the negative frequency operator products to the positive frequency ones. At the same time the operators in observables become normal ordered and the zero-point energy does not appear. Only the symmetrized under the field operators Lagrangians lead to the zero-point energy. The confrontation by the experiments of the such C-symmetric quantization of fields and the solution some of the vacuum energy problems are discussed.

___________________________________________________________

[*)] E-Mail: ctpa@theorphys.org   Web: http://ctpa.theorphys.org          CTPA-07-02

# 1    INTRODUCTION

The spectrum of a non-relativistic particle on the oscillatory potential contains a zero-point energy and this is required by the uncertainty relations and the experiments. The theory describes these physical facts by using a *real valued* coordinate *q* and a *symmetry* of the Hamiltonian under the canonically conjugated variables *q* and *p*. The canonical transformation to the raising and lowering (ladder) operators preserves this symmetry and leads to a symmetrized product of the ladder operators. Then, after the normal arrangement of the operators in the symmetrized product, there appears the zero-point energy:

$$H = \frac{1}{2}(p^2 + \omega^2 q^2) = \frac{1}{2}\omega(a^*a + aa^*) = \omega\left(a^*a + \frac{1}{2}\right). \tag{1}$$

However, if we consider a more general case of oscillating systems where a generalized coordinate should *not necessarily be real valued*, such as relativistic fields containing the negative frequency modes also, then there appears another hermitean quantum Hamiltonian:

$$\begin{aligned} H_+ &= p^*p + \omega^2 q^*q = \omega\left(a_\omega^* a_\omega + a_{-\omega}^* a_{-\omega}\right), \\ q &= \frac{1}{\sqrt{2\omega}}(a_\omega + a_{-\omega}^*), \quad q^* = \frac{1}{\sqrt{2\omega}}(a_\omega^* + a_{-\omega}), \\ p &= \frac{-i\omega}{\sqrt{2\omega}}(a_\omega - a_{-\omega}^*), \quad p^* = \frac{i\omega}{\sqrt{2\omega}}(a_\omega^* - a_{-\omega}). \end{aligned} \tag{2}$$

This Hamiltonian is *asymmetrical* under the complex conjugated variables and for both kind of modes it does not contain the zero-point energy directly.

Thus, the theory of quantum oscillators in fact describes two kind of oscillations - with and without the zero-point energy, and for the relativistic fields the existence or inexistence of *the zero-point vacuum energy* (ZPVE) should be determined experimentally. The lack of the experimental evidences for the ZPVE has been shown recently by R.L. Jaffe [6] who "have presented an argument that the experimental confirmation of the Casimir effect does not establish the reality of zero point fluctuations… Casimir forces can be calculated without reference to the vacuum and, like any other dynamical effect in QED, vanish as α→0… Certainly there is no experimental evidence for the "reality" of zero point energies in quantum field theory (without gravity)… Perhaps there is a consistent formulation of relativistic quantum mechanics in which zero point energies never appear. I doubt it."

In the previous paper [8] it has been proposed the such formulation of quantum field theory without ZPVE where the charge-conjugation symmetry conditions lead to the automatically normal ordered operator products for the observables. In the present paper this method of the *C*-symmetric quantization of fields will be derived in a more simple form and its consequences for the treatments of the vacuum energy will be discussed.

## 2  THE *C*-SYMMETRICAL QUANTIZATION OF BOSON FIELDS

A complex scalar field $\varphi = (\varphi_1 + i\varphi_2)$ has been used conventionally for the description of the electrically charged particles only. But we can use a complex field for the description of the neutral fields also, considering them as a special case of the charged ones at the additional *charge vanishing condition* $Q = 0$. Then, the Lagrangians of the charged and neutral fields have the same form (the neutral fields $\varphi_{(0)}$ are normalized as: $\varphi = \varphi_{(0)}/\sqrt{2}$):

$$L(x) = (\partial_\mu \varphi^*)(\partial^\mu \varphi) - m^2 \varphi^* \varphi. \tag{3}$$

The corresponding Hamiltonian and the charge operator have the form:

$$\begin{aligned} H &= \int d^3 x \left[ (\partial_t \varphi^*)(\partial_t \varphi) + (\nabla \varphi^*)(\nabla \varphi) + m^2 \varphi^* \varphi \right], \\ Q &= i \int d^3 x \left[ \varphi^* (\partial_t \varphi) - (\partial_t \varphi^*) \varphi \right]. \end{aligned} \tag{4}$$

The decomposition of the fields on plane waves with the positive and negative frequency states gives:

$$\begin{aligned} \varphi(x) &= \sum_{\mathbf{k}} \left[ a(k) e^{-ikx} + a(-k) e^{ikx} \right], \\ \varphi^*(x) &= \sum_{\mathbf{k}} \left[ a^*(k) e^{ikx} + a^*(-k) e^{-ikx} \right]. \end{aligned} \tag{5}$$

At the quantization, the non-zero commutators for the creation-annihilation operators of the same frequency sign operators have the form ($\omega \equiv \omega_k = \sqrt{\mathbf{k}^2 + m^2}$):

$$\begin{aligned} \left[ a(k), a^*(k') \right] &= 2\omega (2\pi)^3 \delta^3(\mathbf{k} - \mathbf{k}'), \\ \left[ a(-k), a^*(-k') \right] &= -2\omega (2\pi)^3 \delta^3(\mathbf{k} - \mathbf{k}'), \end{aligned} \tag{6}$$

Therefore, by inserting (5) into (4), $H$ and $Q$ can be represented as:

$$\begin{aligned} H &= \sum_{\mathbf{k}} \omega \left[ a^*(k) a(k) + a^*(-k) a(-k) \right], \\ Q &= \sum_{\mathbf{k}} \left[ a^*(k) a(k) - a^*(-k) a(-k) \right]. \end{aligned} \tag{7}$$

If, as it is accepted in the standard QFT, one *postulates* a direct replacement of the negative frequency operators to the operators of antiparticles,:

$$a(-k) = b^*(k), \quad a^*(-k) = b(k), \tag{8}$$

then there appear ZPVE $H_{(0)}$ and the zero-point charge $Q_{(0)}$ of vacuum [1,2]. However, as it was shown in [8], the such replacement of the operators in $H$ and $Q$ leads to *the breaking of the charge conjugation symmetry* (*C*-symmetry) in the theory, and, therefore, is inadmissible.

Instead of this conventional procedure, in [8] it was proposed a new consistent method of the *C*- and *CP*-symmetric quantization of fields based on the fact that at the quantization of fields, due to the non-linear character of the time reversal, the creation-annihilation operators for the negative frequency modes should be replaced to the operators of antiparticles not directly in the field operators, but in the operator products. However, instead of the complicated time reversal operation, we can use simpler *CP*-symmetry conditions as the natural *constraints* for the *operator products*, and for the *P*-symmetric fields we can simply use only the C-symmetry conditions.

Most simple case is the neutral boson fields where the charge vanishing condition, having the form:

$$Q = \frac{1}{2}\sum_{\mathbf{k}}\left[a_{(0)}^*(k)a_{(0)}(k) - a_{(0)}^*(-k)a_{(0)}(-k)\right] = 0, \qquad (9)$$

directly gives a simple identity for the operator products:

$$a_{(0)}^*(k)a_{(0)}(k) = a_{(0)}^*(-k)a_{(0)}(-k). \qquad (10)$$

This identity allows one to exclude from the Hamiltonian (7) the negative frequency operator product for the neutral fields, and, as the result, to obtain the automatically normal ordered free Hamiltonian without ZPVE:

$$H = \sum_{\mathbf{k}} \omega\, a_{(0)}^*(k)a_{(0)}(k). \qquad (11)$$

In the case of the charged fields the charge-conjugation operation $C$ transforms the operators of the particles to the operators of the charge-conjugated particles $b(\pm k), b^*(\pm k)$ *of the same sign on energy* (up to the inessential phase factors):

$$\begin{aligned}Ca(k)C^{-1} &= b(k), & Ca^*(k)C^{-1} &= b^*(k),\\ Ca(-k)C^{-1} &= b(-k), & Ca^*(-k)C^{-1} &= b^*(-k).\end{aligned} \qquad (12)$$

The charge conjugation of the Hamiltonian $H$ and the charge operator $Q$, however, gives not their former expressions, but the new operators $H_C$ and $Q_C$, expressed through the operators of the charge-conjugated particles of both signs on frequency as:

$$\begin{aligned}H_C &\equiv C H C^{-1} = \sum_{\mathbf{k}} \omega \left[b^*(k)b(k) + b^*(-k)b(-k)\right],\\ Q_C &\equiv C Q C^{-1} = \sum_{\mathbf{k}} \left[b^*(k)b(k) - b^*(-k)b(-k)\right].\end{aligned} \qquad (13)$$

Then one can take into account that the Hamiltonian of the *C*-symmetric fields does not vary at this transformation, while the charge operator, obviously, changes a sign:

$$H = H_C, \quad Q = -Q_C. \qquad (14)$$

Due to (7) and (13), the relations (14) are in fact the constraints for the operator products:

$$a^*(k)a(k)+a^*(-k)a(-k)=b^*(k)b(k)+b^*(-k)b(-k),$$
$$a^*(k)a(k)-a^*(-k)a(-k)=-b^*(k)b(k)+b^*(-k)b(-k). \quad (15)$$

By adding and subtracting these two equalities, one finds the operator *identities*:

$$a^*(k)a(k)=b^*(-k)b(-k), \quad a^*(-k)a(-k)=b^*(k)b(k). \quad (16)$$

Using these identities, one obtains the final expressions for *H* and *Q* of the charged scalar field containing the operators for the *positive-frequency* particles and antiparticles only:

$$H = \sum_{\mathbf{k}} \omega \left[ a^*(k)a(k)+b^*(k)b(k) \right],$$
$$Q = \sum_{\mathbf{k}} \left[ a^*(k)a(k)-b^*(k)b(k) \right]. \quad (17)$$

Thus, due to the *C*-symmetry requirements, *the operators of the antiparticles appear in the normal-ordered form without additional hypotheses*, and ZPVE and the zero-point charge of vacuum do not arise.

### 3   THE *C*-SYMMETRICAL QUANTIZATION OF SPINOR FIELDS

Let's consider the asymmetrical (under the complex conjugated field operators) expressions for the Hamiltonian and the charge operator of a spinor field [1,2]:

$$H = \frac{1}{2}\int d^3x \left[ \psi^+ i\partial_t \psi - (i\partial_t \psi^+)\psi \right], \quad Q = \int d^3x \psi^+ \psi. \quad (18)$$

The momentum decompositions of the field operators gives:

$$\psi(\mathbf{x}) = 2m \sum_{\mathbf{k},\alpha} \left[ b_\alpha(k) u_\alpha e^{-ikx} + b_\alpha(-k) v_\alpha e^{ikx} \right],$$
$$\psi^+(\mathbf{x}) = 2m \sum_{\mathbf{k},\alpha} \left[ b_\alpha^+(k) u_\alpha^+ e^{ikx} + b_\alpha^+(-k) v_\alpha^+ e^{-ikx} \right]. \quad (19)$$

The corresponding non-zero anticommutation relations are:

$$\{b_\alpha(\pm k), b_{\alpha'}^+(\pm k')\} = \frac{\omega}{m}(2\pi)^3 \delta^3(\mathbf{k}-\mathbf{k}')\delta_{\alpha\alpha'}, \quad (20)$$

The substitution of the momentum decomposition of fields then gives:

$$H = 2m\sum_{\mathbf{k},\alpha} \omega \left[ b_\alpha^+(k)b_\alpha(k) - b_\alpha^+(-k)b_\alpha(-k) \right],$$
$$Q = 2m\sum_{\mathbf{k},\alpha} \left[ b_\alpha^+(k)b_\alpha(k) + b_\alpha^+(-k)b_\alpha(-k) \right]. \quad (21)$$

If one accepts the usual *postulate* about the direct replacement of the negative-frequency operators in the form $b_\alpha(-k) = d_\alpha^+(k)$, and $b_\alpha^+(-k) = d_\alpha(k)$, then there appear the infinite ZPVE and the zero-point charge [1,2], and the *C*-symmetry will be broken [8]. Then, here one needs to the *postulate* about the normal ordering.

However, instead of these two compensating each other contradictory procedures, we can simply use the *C*-symmetry conditions for the observables (14) and then, as in the case of the boson fields, we obtain the required *operator identities* for the fermionic operators [8]:

$$b_\alpha^+(k)b_\alpha(k) = -d_\alpha^+(-k)d_\alpha(-k),$$
$$b_\alpha^+(-k)b_\alpha(-k) = -d_\alpha^+(k)d_\alpha(k). \qquad (22)$$

As the result, *H* and *Q* in (21) contain only the positive-frequency operators without ZPVE:

$$H = 2m\sum_{\mathbf{k}}\omega\left[b_\alpha^+(k)b_\alpha(k) + d_\alpha^+(k)d_\alpha(k)\right],$$
$$Q = 2m\sum_{\mathbf{k}}\left[b_\alpha^+(k)b_\alpha(k) - d_\alpha^+(k)d_\alpha(k)\right]. \qquad (23)$$

In the case of the chiral fields it is necessary to use the *CP*-symmetry conditions [8].

In the conventional treatments the vacuum expectation value (v.e.v.) of the fermionic current operator does not vanish [1,2]: $\langle 0|j^\mu(x)|0\rangle = \langle 0|\bar\psi(x)\gamma^\mu\psi(x)|0\rangle \neq 0$ and here the additional anti-symmetrization is required. However, it can be easily shown that in the *C*-symmetrical quantization v.e.v. of the current vanishes automatically without antisymmetrization [8]: $\langle 0|j^\mu(x)|0\rangle = \langle 0|\bar\psi(x)\gamma^\mu\psi(x)|0\rangle = 0$.

## 4 THE LACK OF THE ZERO POINT VACUUM ENERGY AND ITS EXPERIMENTAL CONSEQUENCES

In quantum field theory ZPVE is playing a special role being one of its most contradictory predictions [1-3]. It has been announced, that ZPVE is really detected in form of the Lamb shift and Casimir effect, and that these effects are the experimental proofs of the existence of that form of energy. At the same time, these two effects have been described exactly by other, real physical mechanisms, based on loop diagrams for field's quanta or fluctuating fields of sets of vibrating atoms [1-6].

Notice, that in the literature these two treatments have been represented as *the alternative methods of description of the same phenomena*. In fact, an origin of the such dual treatment is that the notion "*vacuum fluctuations*" has been used in the literature in the *various senses*. Thus, using this notion in one of senses, then the obtained results have been assigned to another sense of this notion and the observation of the effects of one type of vacuum fluctuations have been treated as a "proof" of the existence in principle different type of fluctuations. Therefore, one must to clearly understand in what of the senses this notion has been used in every case.

*Firstly* the "vacuum fluctuations" means the processes of creation and annihilation of the virtual quanta of relativistic fields described by the loop diagrams, the consequences of which are experimentally well checked. They are *the fluctuations of the physical vacuum* of the quantized fields, determined mainly by the interaction Hamiltonian $H_I$, and the such loop fluctuations do not have any relation to ZPVE in the free Hamiltonian $H_0$. Nevertheless, as in the case of the Lamb shift, the action of these fluctuations to the electrons in atoms may be *imitated* by the small vibrations of electrons in atoms due to some *effective* oscillatory potential, which, of course, has the zero-point energy. The such phenomenological approach (Welton, 1948) is well known [1,2], but these zero-point vibrations of atoms in the effective oscillatory potential approximating the complicated loop contributions have not any relation to the vibrations of electrons due to the external vacuum fields representing the ZPVE of the *free* Hamiltonian.

*The second sense* of the notion «vacuum fluctuations» concerns the fluctuating external fields of real *vibrating sources*, particularly, atoms in solid states. Vibrating atoms in crystals can be represented as a set of harmonic oscillators (containing the zero-point energies), and the electromagnetic fields of the such atoms, due to the overlapping, form an effective fluctuating field of the crystal leading to many observable effects [4-6]. Since they are exist at *zero temperature* of the body, they also have been named as *the zero-point fluctuations.* This fluctuating electromagnetic field of the solid states again has no any relation to the zero-point fluctuations of the free electromagnetic field's vacuum.

And, at last, *in the third sense* the notion "vacuum fluctuations" means the hypothetical *zero-point fluctuations of vacuum fields* generating the ZPVE in the *free* Hamiltonians of fields $H_0$. It has been supposed that if there is some energy of field's oscillations, therefore, there are corresponding fluctuating pure (external) vacuum fields. In the case of electromagnetic field the such fluctuating external vacuum fields $\mathbf{E}_{(0)}, \mathbf{H}_{(0)}$ have been introduced into the free Hamiltonian $H_0$ instead of the ZPVE as [1-3]:

$$H_0 = \sum_{\mathbf{k},\lambda} \omega\, a_\lambda^* a_\lambda + H_{(0)}, \qquad H_{(0)} = 2V \int \frac{d^3 k}{(2\pi)^3} \frac{\omega_k}{2} = \int_V d^3 x \frac{1}{2}\left(\mathbf{E}_{(0)}^2 + \mathbf{H}_{(0)}^2\right), \qquad (24)$$

where $V$ is a normalization volume. Notice again, that these fluctuations have no relation neither to quanta of fields, nor to sources, real or virtual, and represent a new physical hypothesis *in addition* to the observable physical effects of the quanta or the sources.

The existence of the first two kinds of fluctuations are the experimental fact and their contributions to the observable effects obviously must be taken into account. However, historically the results *of the approximate calculations* of the first two kinds of the real fluctuations of the fields (the fluctuations of the physical vacuum and the fields of the sources) *have been misinterpreted* as the observation of the third kind of fluctuations - the zero-point fluctuations of the pure vacuum fields. As the result, two mistaken conclusions have been done:

first, that the description of the effects in terms of the fluctuations of the physical vacuum (first two types of fluctuations) or in terms of the zero-point fluctuations of the vacuum fields (the third type of fluctuations) are two equivalent descriptions of the same phenomena, and second, that the observations of these effects confirm the existence of the zero-point fluctuations of the vacuum fields.

At last, notice that three different contributions to the *cosmological constant* $\Lambda$ from the vacuum energy density of the quantized fields have been widely discussed in the literature [7]. They are ZPVE, the vacuum loop diagrams and the energy of the condensates at spontaneous symmetry breaking. The last two contribution to $\Lambda$ (loop diagrams and condensates) are not the subject of the present paper. However, the gravitational effect of the first one, ZPVE, the contribution of which to $\Lambda$ in the conventional treatments diverges, was one of main difficulties of quantum field theory. The proven exact vanishing of ZPVE in quantum field theory at strictly following to the C-symmetry requirements and appropriately interpreting the experiments leads to the lack of the contribution of ZPVE to $\Lambda$ and this fact *partly* solves the cosmological constant problem.

## 5    CONCLUSION

Thus, the asymmetrical under the complex conjugated field operators Lagrangians do not lead to ZPVE and the zero-point charge due to the *C*-symmetry conditions, and only the symmetrical under the fields Lagrangians of the relativistic fields lead to the divergent ZPVE.

The observable effects correspond or to the fluctuations of the physical vacuum describing by the loop diagrams (Lamb shift), or to the fluctuations of fields of the real sources (Casimir effect) and they do not relate to ZPVE. Therefore, the experiments with high accuracy exclude the existence of the ZPVE of the relativistic fields and allow one to use only the Lagrangians asymmetrical under the complex conjugated field operators.

Since ZPVE for the relativistic fields exactly vanishes, it does not contribute to the cosmological constant.